# Analog-to-Digital Converter Based on Voltage-controlled Superconducting Device

Md Mazharul Islam[1], Connor A. Good[2], Diego Ferrer[1], Juan P. Mendez[3], Denis Mamaluy[3], Wei Pan[4], Kathleen E Hamilton[5], and Ahmedullah Aziz[1*]

[1] Dept. of Electrical Eng. and Computer Sci., University of Tennessee, Knoxville, TN, 37996, USA
[2] Dept. of Computer Eng. University of Mount Union, Alliance, OH 44601, USA
[3] Sandia National Laboratories, Albuquerque, NM 87123, USA
[4] Sandia National Laboratories, Livermore, CA 94550, USA
[5] Oak Ridge National Laboratory, Oak Ridge, TN 37831, USA
[*]Corresponding Author. Email: aziz@utk.edu

*Abstract-* **The increasing demand for cryogenic electronics in superconducting and quantum computing systems calls for ultra–energy-efficient data conversion architectures that remain functional at deep cryogenic temperatures. In this work, we present the first design of a voltage-controlled superconducting flash analog-to-digital converter (ADC) based on a novel quantum-enhanced Josephson junction field-effect transistor (JJFET). Exploiting its strong gate tunability and transistor-like behavior, the JJFET offers a scalable alternative to conventional current-controlled superconducting devices while aligning naturally with CMOS-style design methodologies. Building on our previously developed Verilog-A compact model calibrated to experimental data, we design and simulate a three-bit JJFET-based flash ADC. The core comparator block is realized through careful bias current selection and augmented with a three-terminal nanocryotron to precisely define reference voltages. Cascaded JJFET comparators ensure robust voltage gain, cascadability, and logic-level restoration across stages. Simulation results demonstrate accurate quantization behavior with ultra-low power dissipation, underscoring the feasibility of voltage-driven superconducting mixed-signal circuits. This work establishes a critical step toward unifying superconducting logic and data conversion, paving the way for scalable cryogenic architectures in quantum–classical co-processors, low-power AI accelerators, and next-generation energy-constrained computing platforms.**

*Index Terms-* **Josephson junction, Field effect transistor, Cryogenic, Logic,**

The rapid expansion of quantum computing processors, superconducting accelerators, and cryogenic AI hardware has created an urgent need for energy-efficient mixed-signal circuits that can reliably operate at deep cryogenic temperatures [1]–[3]. Among these, analog-to-digital converters (ADCs) are essential, serving as the critical interface between the analog domain of quantum measurements and sensor front-ends and the digital domain of control and processing backends [4]–[6]. However, conventional CMOS-based ADCs face severe limitations in this regime. In large-scale qubit systems, the reliance on CMOS technology leads to massive wiring overhead, large footprint, and heat dissipation, ultimately limiting overall integration capability and scalability [7], [8]. Moreover, CMOS devices suffer from excessive power consumption [9], degraded characteristics [10], limited fidelity [11], and reduced functionality [12] under cryogenic biasing conditions, posing a significant barrier to the development of large-scale cryogenic systems.

Superconducting electronics provides a compelling alternative, offering ultra-low switching energy (down to the attojoule scale), ultra-high bandwidths reaching hundreds of gigahertz, and intrinsic compatibility with cryogenic environments [8], [13]–[17]. To date, however, most superconducting logic and mixed-signal circuits have relied on current-controlled single-flux quantum (SFQ) Josephson junction device [18]–[20]. While these technologies have enabled exceptional performance, these devices require substantial DC bias currents, which in large-scale systems can generate unwanted stray magnetic fields, heat, and operational instabilities [21]–[23]. Moreover, their limited fan-out capability and lack of voltage gain fundamentally restrict scalability, especially for system-level functions such as data conversion[24], [25].

Recent progress in voltage-controlled superconducting devices, particularly the Josephson junction field-effect transistor (JJFET), offers a promising pathway toward CMOS-like, voltage-driven design methodologies at cryogenic temperatures. In our prior work, we introduced a Verilog-A-based compact model [26] of the quantum-enhanced JJFET [27], leveraging an InAs/GaSb heterostructure that supports an excitonic insulator phase transition [28], and demonstrated its ability to realize fundamental Boolean logic gates and multistage logic circuits [26]. These results established the JJFET as a viable building block for scalable, voltage-controlled superconducting logic.

In this work, we build on this foundation and extend our exploration in mixed-signal circuit design by proposing the first ADC architecture based on JJFETs. We design voltage comparator block based on JJFET and a three-terminal electrothermal superconducting switching element known as nanocryotron (nTron) [29]. Using the experimentally calibrated compact model and prior Boolean logic gate topologies [26], we design and simulate a multi-bit flash ADC [30], demonstrating evenly spaced quantization behavior with low power dissipation under cryogenic operating conditions. By unifying superconducting logic and data conversion within the JJFET platform, this work introduces a scalable pathway toward cryogenic mixed-signal architectures, enabling seamless integration of superconducting circuits with quantum processors, low-power cryogenic AI accelerators, and hybrid classical–quantum systems.

At the device level, the JJFET offers a voltage-controlled pathway to modulate supercurrent in a weak-link channel, serving as a superconducting equivalent of the MOSFET (Fig 1a). In its quantum-enhanced form, the JJFET is realized using an InAs/GaSb zero-gap heterostructure [28] contacted by superconducting electrodes. This material platform enables a gate-tunable excitonic insulator (EI) phase transition, where electron–hole pair condensation sharply modifies the carrier coherence length. The resulting strong dependence of the Josephson critical current ($I_{c\_JJ}$) on gate voltage provides the JJFET with enhanced gain and transistor-like behavior. In practice, the device exhibits two distinct resistance states: sub-gap resistance ($R_{SG}$) and normal-state resistance ($R_N$) with a gate-controlled transition governed by $I_{c\_JJ}$. This highly nonlinear response enables the JJFET to emulate digital ON/OFF states under appropriate biasing, making it an attractive candidate for circuit-level integration.

To facilitate the design and simulation based on the quantum-enhanced JJFET, we previously developed a Verilog-A-based compact model of the JJFET [26]. This model incorporates the experimentally observed gate dependence of both the critical current and channel resistances, enabling accurate reproduction of the nonlinear device characteristics. Specifically, the model employs a phenomenological resistively capacitance shunted junction (RCSJ) model, along with a piecewise-linear fit for $I_c$, calibrated against experimental data from the InAs/GaSb heterostructures [27]. Comparison with experimental I–V measurements confirms that the model accurately reproduces the gate-tunable resistance modulation, and the transition between superconducting and resistive regimes (Fig. 1b). By embedding this model into the industry-standard circuit simulator, HSPICE, we enable the design and analysis of JJFET-based logic circuit. While the JJFET provides the essential gate-controlled switching mechanism, voltage gain and fan-out restoration remain critical for cascadability and multistage logic design. To address this, we combine the JJFET with the three-terminal nTron. The nTron operates by injecting a small gate current into a nanowire choke region, inducing a localized resistive hotspot that suppresses superconductivity in the adjacent channel. This mechanism enables a small control signal to modulate a larger current, effectively providing digital gain and voltage restoration. By integrating JJFETs with nTrons, we demonstrated the successful realization of a family of Boolean logic gates, including NOT, NAND, NOR, and a three-input majority gate [26] (Fig. 1(d-i)). These gates exhibit correct functionality, cascadability, and compatibility with voltage-based signaling conventions. Collectively, the JJFET–nTron framework establishes a foundation for voltage-controlled superconducting logic that mirrors CMOS-style methodologies. Together, these devices and logic blocks can be readily employed in broader voltage-driven superconducting systems, including mixed-signal architectures such as the flash ADC demonstrated in this work.

The basic building block of a flash ADC requires comparator blocks with variable reference voltages. In a flash ADC, the analog voltage is converted to a thermometer code in the comparator stage depending on the level of input voltage. Fig. 2 (a, b, c) illustrates our proposed topology of the JJFET-based comparator block. The JJFET channel resistance decreases with the increase of the input voltage (Fig. 2e). Here, the applied current $I_{Bias1}$ is divided between the JJFET channel and the parallel resistance ($R_P$). An nTron is connected in series with the channel of the JJFET. When the nTron gate current ($I_g$) is greater than the critical current ($I_c$) of the nTron channel, the nTron channel turns resistive. Meanwhile, the nTron channel is biased by a channel current ($I_{Bias2}$) and as the nTron channel turns resistive, a voltage is dissipated across the nTron channel. With an appropriate selection of $V_{bias}$ (-0.7 V), it is possible to produce the necessary voltage at the output to drive the next stage of logic gates. To set the reference voltage of our proposed comparator block topology, $I_{Bias1}$ can be tuned so that the nTron channel switches at a particular value of the input voltage level. This way, we can tune the reference voltage of the comparator block. The conditions for comparator output can be described by the following set of equations.

$$I_g = \frac{R_P}{R_P + R_{JJ}(f(V_{IN})) + R_S} \times I_{Bias1}$$

$$V_{OUT} = \begin{cases} (I_{Bias2} \times R_{nTron}) + V_{bias} & \text{if } V_{IN} > V_{ref} \\ V_{bias} & \text{if } V_{IN} < V_{ref} \end{cases} \quad (1)$$

when $V_{ref} = V_{IN}, I_g = I_c$ (Fig. 2(d)), and we choose in our design

$$V_{OUT} = \begin{cases} 0 \text{ V} & ; \text{if } V_{IN} > V_{ref} \\ -0.7 \text{ V} & ; \text{if } V_{IN} < V_{ref} \end{cases}$$

By increasing the $I_{Bias1}$, we can decrease the input voltage ($V_{IN}$) at which the nTron channel turns resistive and we get a high voltage (0 V) at the output representing logic '1' (Figs. 2(d,e)). However, there is a discontinuity of the JJFET-channel resistance with $V_g$ as shown in Fig. 2(e). This discontinuity can be attributed to the inherent nonlinearity or experimentally observed discontinuity in the switching dynamic of the device. Future device improvements might attenuate this discontinuous behavior. However, to address this current issue, we avoid setting $V_{ref}$ within this region. Instead, we utilize the simple voltage division rule by employing two series resistances as shown in Fig. 2(f). Here, $V_{IN}$ is taken from $R_2$ instead of being directly applied at the gate of the JJFET. If, we choose $R_1 = R_2$, then $I_{Bias1}$ can be chosen for $\frac{V_{ref}}{2}$, instead of $V_{ref}$. For example, if we want a comparator block with $V_{ref}$ = -0.45 V, we can choose corresponding $I_{Bias1}$ for $V_{ref}$ = -0.225 V and avoid the discontinuous region.

Fig. 3(a) illustrates the basic building block of our proposed 3-bit flash ADC architecture. At the beginning of the architecture, there is a comparator stage that converts analog voltage levels into 7-bit thermometer code. For an n-bit ADC, $2^n - 1$ bit thermometer code is generated that creates $2^n$ distinct levels of output. This thermometer code is fed into a priority encoder block to create n-bit binary output. In our proposed 3-bit ADC, we need seven comparator blocks with evenly spaced reference voltages (Fig. 3). These reference voltages were chosen by precisely selecting $I_{Bias1}$(Fig.2(d)). Fig. 3(c) illustrates the output of the comparator stage for an applied ramp (-0.7 V to 0 V). From the figure, we can see the generation of evenly spaced eight distinct thermometer codes from 7'b0000000 to 7'b1111111 (Fig. 3(b)). This design is readily scalable for higher number of bits where each bit of thermometer code will require one comparator with appropriate selection of $I_{Bias1}$ as described in equation 1. The conversion of thermometer to binary code can be done by implementing the Boolean expressions:

$$\begin{aligned}
D1 + D3 + D5 + D7 &= \text{Bit 0} \\
D2 + D3 + D6 + D7 &= \text{Bit 1} \\
D4 + D5 + D6 + D7 &= \text{Bit 2} \\
A\,\bar{B}\,\bar{C}\,\bar{D}\,\bar{E}\,\bar{F}\,\bar{G} &= D1 \\
B\,\bar{C}\,\bar{D}\,\bar{E}\,\bar{F}\,\bar{G} &= D2 \\
C\,\bar{D}\,\bar{E}\,\bar{F}\,\bar{G} &= D3 \\
D\,\bar{E}\,\bar{F}\,\bar{G} &= D4 \\
E\,\bar{F}\,\bar{G} &= D5 \\
F\bar{G} &= D6 \\
G &= D7
\end{aligned} \qquad (2)$$

Fig. 4 shows the overall schematics for implementation of the thermometer-to-binary encoder module that implements the Boolean functions expressed in equation (2). Here, we have utilized the JJFET-based NAND, NOR and NOT gate [26]. The precise switching delay for the quantum-enhanced JJFET has not been measured. However, theoretical and experimental works show that Josephson devices can operate at frequencies up to several hundred gigahertz (GHz) and even into the terahertz (THz) range, we can envision that our ADC can successfully operate at the level of hundreds of GHz. Fig. 5 illustrates the simulation waveform of our proposed quantum-enhanced JJFET-based 3-bit flash ADC. The power consumption for our proposed ADC is estimated to be ~823 $\mu$W. Table I summarizes the comparison of our proposed ADC with several state-of-the-art cryogenic CMOS-based ADC topologies [31]–[35]. From the table, it is evident that our proposed ADC topology works at a power which is orders of magnitude lower than most of the state-of-the-art CMOS-based cryogenic ADC. We also emphasize that further optimization of the quantum-enhanced JJFET [27], which is still at the research stage, will lead to lower power consumption.

We further evaluated the nonlinearity of the proposed ADC by analyzing its quantized output levels. The differential nonlinearity (DNL) measures the deviation of each output code width from the ideal least significant bit (LSB), thereby indicating the uniformity of quantization steps. The integral nonlinearity (INL) represents the cumulative deviation of the ADC transfer characteristic from a best-fit straight line, reflecting the overall linearity of conversion. These metrics are defined as:

$$DNL(k) = \frac{V_{k+1} - V_k}{V_{LSB,ideal}} - 1$$
$$INL(k) = \sum_{i=0}^{k} DNL(i)$$

Table II summarizes the extracted DNL and INL values for the proposed JJFET-based ADC. The results confirm that the quantized outputs closely follow the ideal transfer characteristic, validating the accuracy of the architecture.

To summarize, we have introduced and demonstrated the first design of a voltage-controlled superconducting flash ADC based on quantum-enhanced Josephson junction field-effect transistors (JJFETs). Leveraging the strong gate tunability and transistor-like behavior of the JJFET, combined with the gain and restoration capabilities of the nTron, we established a scalable comparator framework that enables robust thermometer-to-binary conversion at cryogenic temperatures. The proposed three-bit flash ADC achieves accurate quantization with ultra-low power dissipation, highlighting the promise of non-flux-based, voltage-driven superconducting mixed-signal circuits. Beyond its immediate functionality, this work underscores the broader potential of JJFET-based platforms to unify logic and data conversion within a single superconducting design methodology, offering a CMOS-like paradigm in the cryogenic regime. Compared to state-of-the-art cryogenic CMOS ADCs, our design demonstrates significantly reduced power consumption and compatibility with high-speed operation, making it particularly attractive for integration in quantum–classical co-processors, cryogenic AI accelerators, and other energy-constrained superconducting systems.

Looking ahead, further experimental validation of the JJFET-based flash ADC will be essential, particularly with respect to dynamic switching speed, noise resilience, and scalability to higher-bit resolutions. Moreover, the seamless co-integration of JJFET-based logic and mixed-signal blocks paves the way for a new generation of cryogenic circuit architectures that bridge the gap between quantum devices and classical control electronics. Ultimately, this work provides a critical step toward establishing energy-efficient, scalable, and fully voltage-controlled superconducting systems for next-generation computing platforms.

**Acknowledgement:**
The work at Sandia National Laboratories (SNL) is supported by an LDRD project. DF is partially supported by the SMART program at SNL. SNL is a multi-mission laboratory managed and operated by National Technology & Engineering Solutions of Sandia, LLC (NTESS), a wholly owned subsidiary of Honeywell International Inc., for the U.S. Department of Energy's National Nuclear Security Administration (DOE/NNSA) under contract DE-NA0003525. This written work is authored by an employee of NTESS. The employee, not NTESS, owns the right, title and interest in and to the written work and is responsible for its contents. Any subjective views or opinions that might be expressed in the written work do not necessarily represent the views of the U.S. Government. The publisher acknowledges that the U.S. Government retains a non-exclusive, paid-up, irrevocable, world-wide license to publish or reproduce the published form of this written work or allow others to do so, for U.S. Government purposes. The DOE will provide public access to results of federally sponsored research in accordance with the DOE Public Access Plan.

Analog-to-Digital Converter Based on Voltage-controlled Superconducting Device## Author Contributions

M.M.I. conceived the idea, designed the ADC, and performed the simulations. C.A.G and D.F. performed several simulation works. J. P. M., D. M., W. P. provided insights into the device's characteristics. A.A. supervised the project. All authors commented on the results and wrote the manuscript.

## Competing Interests

The authors declare no competing interests.

## Data Availability

The data that supports the plots within this paper and other findings of this study are available from the corresponding author upon reasonable request.

## References

[1] F. Tacchino, P. Barkoutsos, C. Macchiavello, I. Tavernelli, D. Gerace, and D. Bajoni, "Quantum implementation of an artificial feed-forward neural network," *Quantum Sci. Technol.*, 2020, doi: 10.1088/2058-9565/abb8e4.
[2] J. Chen, "Review on Quantum Communication and Quantum Computation," in *Journal of Physics: Conference Series*, 2021, doi: 10.1088/1742-6596/1865/2/022008.
[3] M. Onen, B. A. Butters, E. Toomey, T. Gokmen, and K. K. Berggren, "Design and Characterization of Superconducting Nanowire-Based Processors for Acceleration of Deep Neural Network Training," *arXiv*, 2019.
[4] J. Lisenfeld, "Superconductor Digital Electronics: Quantum Computing," in *Applied Superconductivity: Handbook on Devices and Applications*, 2015.
[5] G. Kiene, R. W. J. Overwater, A. Catania, A. G. Sreenivasulu, P. Bruschi, E. Charbon, M. Babaie, and F. Sebastiano, "A 1-GS/s 6-8-b Cryo-CMOS SAR ADC for Quantum Computing," *IEEE J. Solid-State Circuits*, 2023, doi: 10.1109/JSSC.2023.3237603.
[6] H. Homulle, S. Visser, and E. Charbon, "A Cryogenic 1 GSa/s, Soft-Core FPGA ADC for Quantum Computing Applications," *IEEE Trans. Circuits Syst. I Regul. Pap.*, 2016, doi: 10.1109/TCSI.2016.2599927.
[7] "Using Cryogenic CMOS Control Electronics to Enable a Two-Qubit Cross-Resonance Gate," *PRX Quantum*, 2024, doi: 10.1103/PRXQuantum.5.010326.
[8] M. Ahmad, C. Giagkoulovits, S. Danilin, M. Weides, and H. Heidari, "Scalable Cryoelectronics for Superconducting Qubit Control and Readout," *Adv. Intell. Syst.*, 2022, doi: 10.1002/aisy.202200079.
[9] N. Zhuldassov and E. G. Friedman, "Cryogenic dynamic logic," in *Proceedings - IEEE International Symposium on Circuits and Systems*, 2020, doi: 10.1109/iscas45731.2020.9181079.
[10] Y. Zhang, J. Xu, T. T. Lu, Y. Zhang, C. Luo, and G. Guo, "Hot Carrier Degradation in MOSFETs at Cryogenic Temperatures down to 4.2 K," *IEEE Trans. Device Mater. Reliab.*, 2021, doi: 10.1109/TDMR.2021.3124417.
[11] B. Patra, "CMOS circuits and systems for cryogenic control of silicon quantum processors," 2021, doi: 10.4233/UUID:CEA59727-FDA2-41E1-BA87-9404EF22202D.
[12] "Cryo-CMOS Circuits and Systems for Quantum Computing Applications," *IEEE J. Solid-State Circuits*, 2018, doi: 10.1109/JSSC.2017.2737549.
[13] S. Alam, M. M. Islam, M. S. Hossain, A. Jaiswal, and A. Aziz, "Cryogenic In-Memory Bit-Serial Addition using Quantum Anomalous Hall Effect-based Majority Logic," *IEEE Access*, pp. 1–1, 2023, doi: 10.1109/ACCESS.2023.3285604.
[14] M. M. Islam, S. Alam, M. S. Hossain, K. Roy, and A. Aziz, "Cryogenic Neuromorphic Hardware," *arXiv Prepr.*, Mar. 2022, doi: 10.48550/arxiv.2204.07503.
[15] S. Alam, M. M. Islam, M. S. Hossain, K. Ni, V. Narayanan, and A. Aziz, "Cryogenic Memory Array based on Ferroelectric SQUID and Heater Cryotron," in *Device Research Conference - Conference Digest, DRC*, 2022, doi: 10.1109/DRC55272.2022.9855813.
[16] M. M. Islam, S. Alam, M. S. Hossain, and A. Aziz, "Dynamically Reconfigurable Cryogenic Spiking Neuron based on Superconducting Memristor," in *IEEE international conference on Nanotechnology*, 2022, pp. 307–310, doi: 10.1109/NANO54668.2022.9928634.
[17] M. M. Islam, S. Alam, C. D. Schuman, M. S. Hossain, and A. Aziz, "A Deep Dive Into the Design Space of a Dynamically Reconfigurable Cryogenic Spiking Neuron," *IEEE Trans. Nanotechnol.*, vol. 22, pp. 666–672, 2023,6

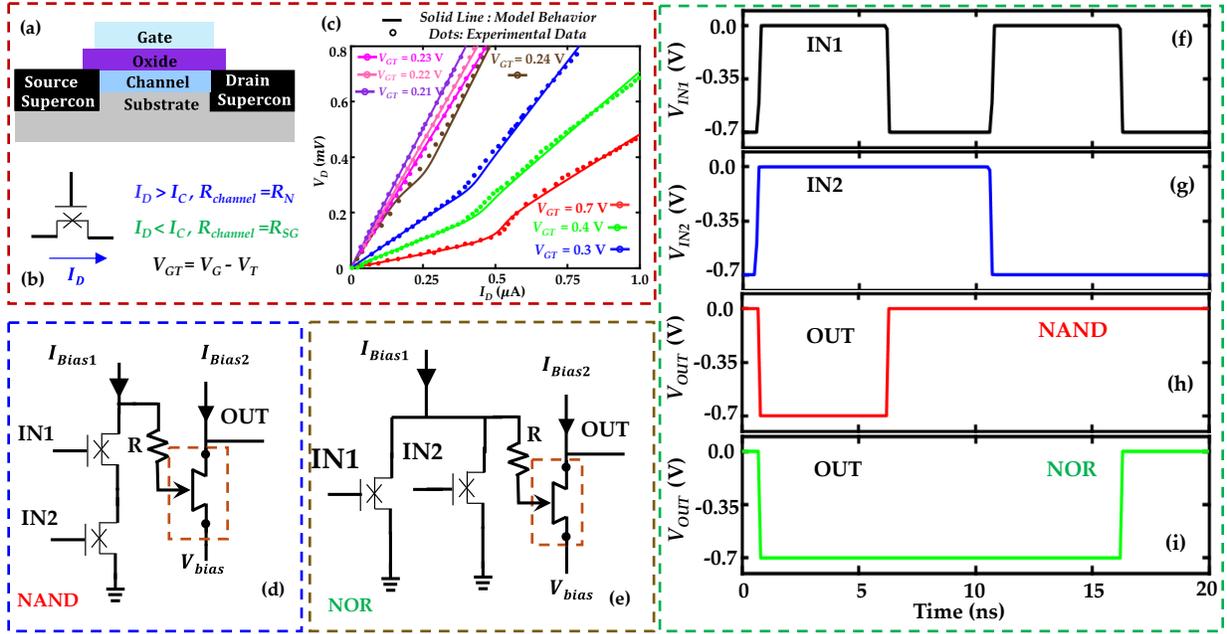

**Fig. 1.(a)** Device structure, and **(b)** circuit symbol of the quantum-enhanced Josephson Junction Field-Effect Transistor (JJFET), where superconducting Tantalum contacts define the source and drain, and the channel consists of a zero-gap InAs/GaSb heterostructure. **(c)** Simulated drain voltage ($V_D$) vs drain current ($I_D$) characteristics for different $V_{GT}$. Here, $V_{GT} = V_G - V_T$, and $V_T = -0.24$V. The compact model characteristics (solid line) are plotted alongside with the experimental data reported in [27] (dotted data). JJFET-based **(d)** NAND and **(e)** NOR gate circuit topology. **(f)-(g)** Applied voltage inputs (IN1 and IN2) and corresponding output (OUT) for **(h)** NAND, and **(i)** NOR gate.

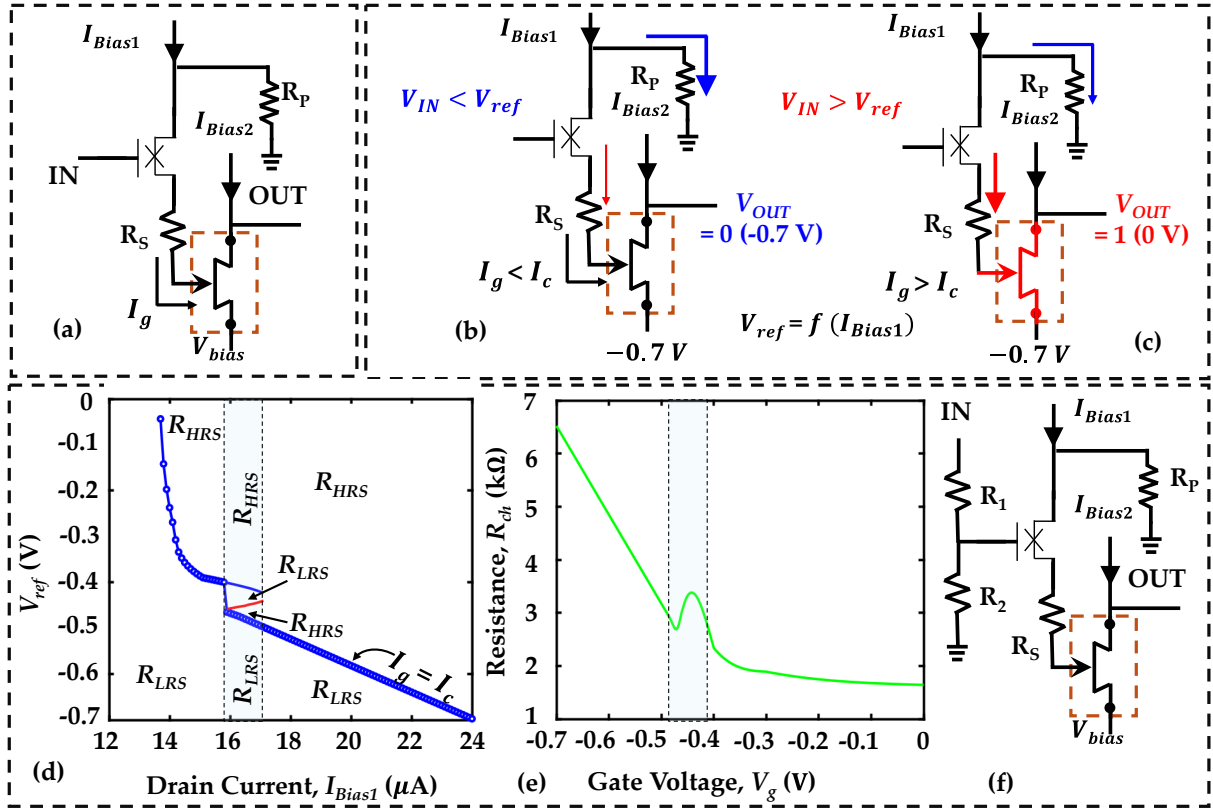

**Fig. 2.(a)** Circuit topology of JJFET-based comparator circuit. Here, the voltage reference is set by accurate selection of $I_{Bias1}$. **(b)** Working principle of the comparator, when $V_{IN} < V_{ref}$. **(c)** Working principle of the comparator, when $V_{IN} > V_{ref}$. **(d)** Reference voltage ($V_{ref}$) selection based on applied $I_{Bias1}$. For a particular $I_{Bias1}$, the comparator JJFET remains in the high resistive state ($R_{HRS}$) below the blue line, and in the low resistive state ($R_{LRS}$) above the blue line. This way, corresponding $I_{Bias1}$ can be selected for a particular $V_{ref}$. **(e)** Channel resistance ($R_{channel}$) variation of the JJFET with applied gate voltage ($V_g$). Within the marked region, $R_{channel}$ exhibits a change of its behavior ($R_{channel}$ increases and then decreases again with the increase of $V_g$). This is marked as the red line in Fig.2(d), which marks as threshold voltage above which, the $R_{channel} = R_{HRS}$ again and $R_{channel} = R_{LRS}$ above the thin blue line. **(f)** Comparator circuit topology for $V_{ref}$ within the marked region illustrated in Figs. 2 (d, e).

Analog-to-Digital Converter Based on Voltage-controlled Superconducting Device

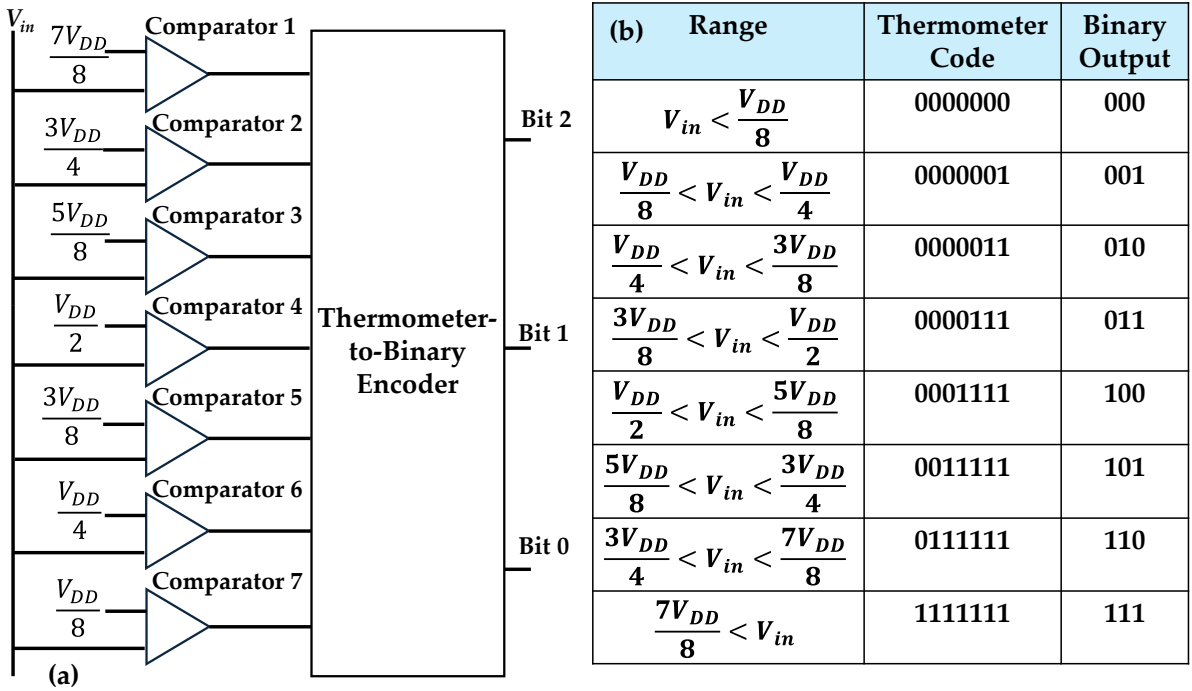

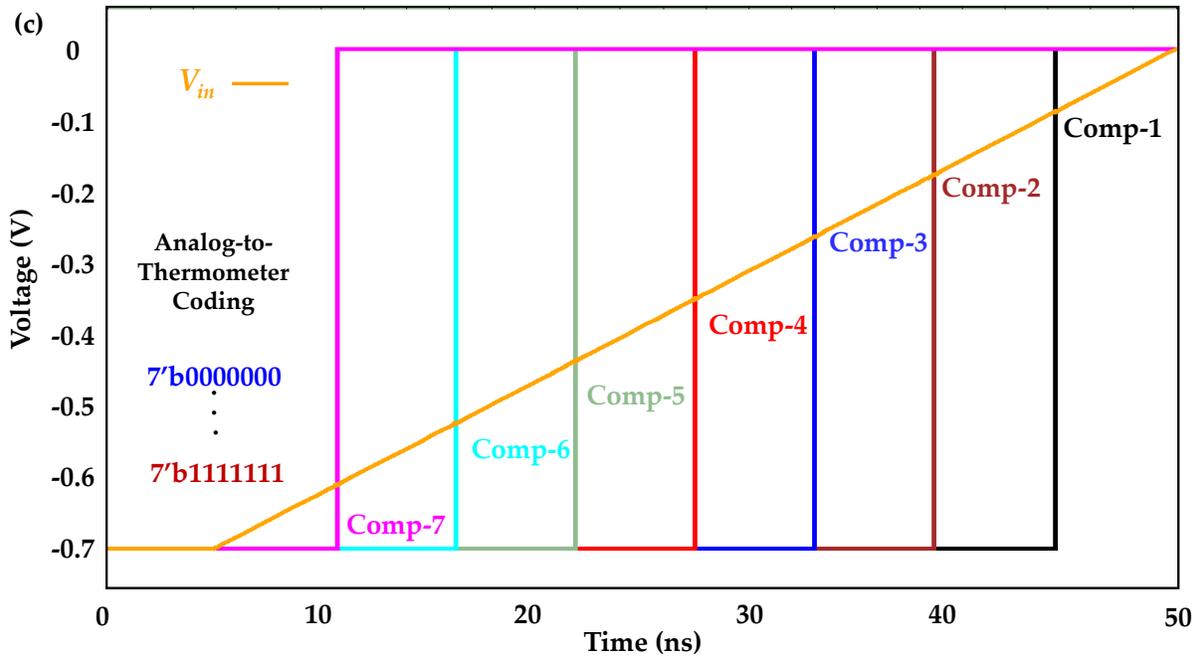

**Fig. 3.(a)** 7-bit binary thermometer code for an applied ramped input voltage. Eight evenly spaced thermometer codes are generated (from 7'b0000000 to 7'b1111111). **(b)** 3-bit flash ADC architecture with comparators and a thermometer-to-binary encoder block. **(c)** Table summarizing eight different voltage ranges and corresponding thermometer and 3-bit binary outputs. Here, $V_{DD}$ = -0.7V.

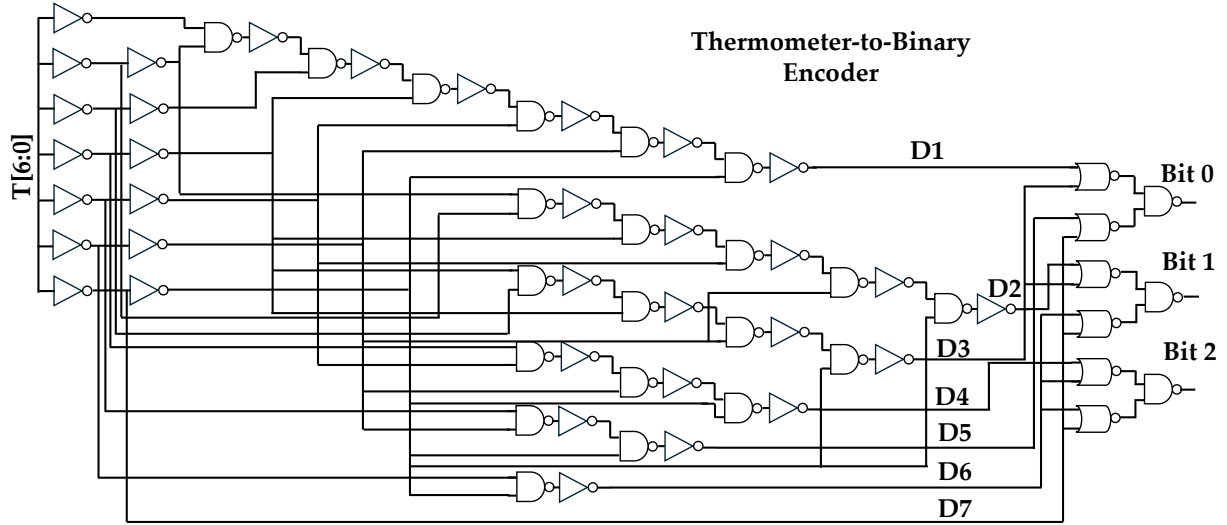

**Fig. 4.** 7-bit thermometer to 3-bit binary encoder for the 3-bit flash ADC utilizing the JJFET-based 2-input NAND, 2-input NOR, and NOT gate. **(b)** 3-bit binary output waveform corresponding to the applied ramp voltage from -0.7V to 0V.

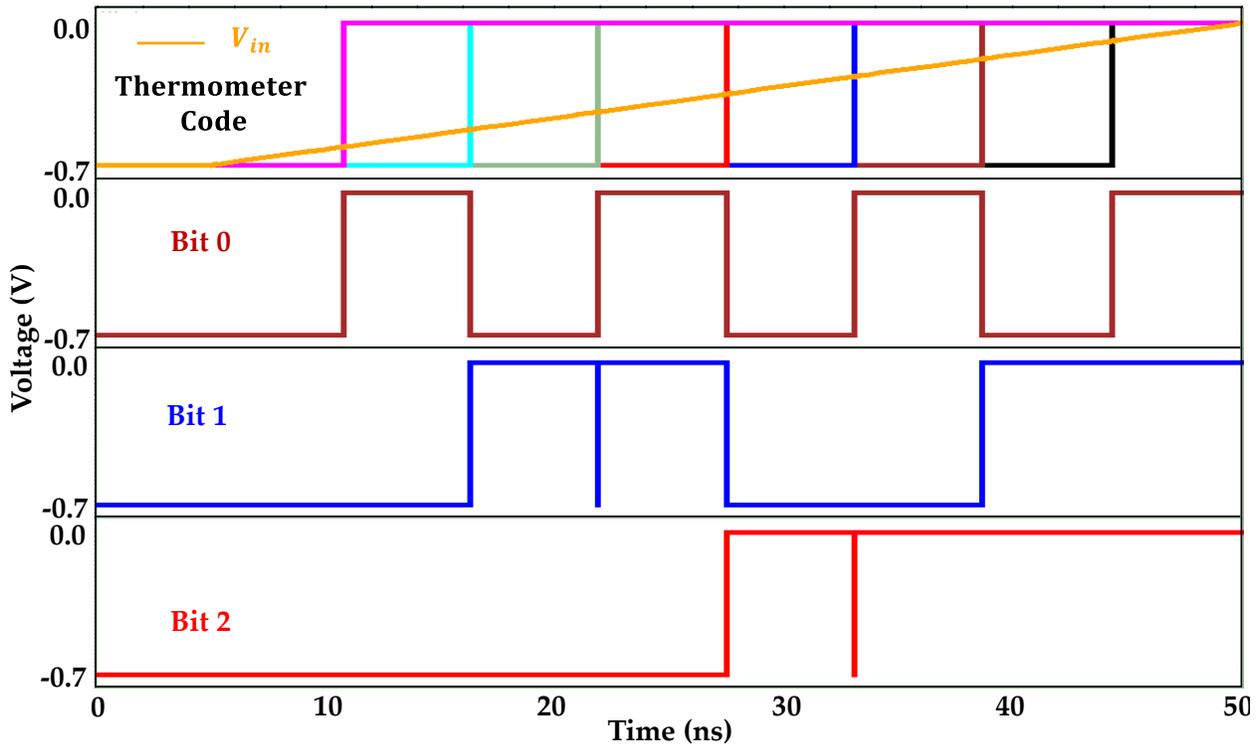

**Fig. 5.** 3-bit binary output waveform corresponding to the applied ramp voltage from -0.7V to 0V.

Table I: Comparison of JJFET-based flash ADC topology with several state-of-the-art ADCs.

| ADC Application | Core Platform | Configuration | Operating Range (V) | Power (W) | Latency/Sampling Rate | Ref. |
|---|---|---|---|---|---|---|
| Superconducting Qubits | Teledyne EV8AQ160 (CMOS) |  | Vcc = 3.15, Vccd = 1.7, Vcco = 1.7 | 4.2 W @ Room Temp | 3ns | [34] |
| Superconducting Qubits | CMOS | FPGA-based | 0.9 - 1.6 V | 750 mW @ 15K | 1200 MS/s | [33] |
| Spin Qubits | CMOS | SAR | Core = 1.1, Clock = 2.5 | 0.813 mW @ 150K | 1000 MS/s (~1ns) | [32] |
| Space Imaging/Particle Experiments | CMOS | SAR | 5V | 350 uW @ 4.2K | 3 kHz | [31] |
| Neutrino Experiments | CMOS | Custom (ColdADC_P2) | 2.25V | 331.8 mW @ 77K | 2 MS/s (~500ns) | [30] |
| -- | JJFET | Flash | -0.7V – 0V | ~823 $\mu$W @ 11 mK | - | This Work |

Table II: Differential nonlinearity (DNL) and integral nonlinearity (INL) of JJFET-based ADC

| ADC Application | Step Width | DNL | INL |
|---|---|---|---|
| 000 → 001 | 0.894444 | 0.02222 | 0.02222 |
| 001 → 010 | 0.862556 | -0.01422 | 0.008 |
| 010 → 011 | 0.870333 | -0.00533 | 0.002667 |
| 011 → 100 | 0.869556 | -0.00622 | -0.00356 |
| 100 → 101 | 0.871111 | -0.00444 | -0.008 |
| 101 → 110 | 0.869556 | -0.00622 | -0.01422 |
| 110 → 111 | 0.880444 | 0.006222 | -0.008 |